\documentclass[prb, twocolumn, showpacs, amsmath, amssymb]{revtex4-1}

\usepackage{amssymb,amsfonts,amsmath} 
\usepackage{graphicx}
\usepackage{color}
\usepackage{hyperref}
\usepackage{longtable}
\newcommand{\ix}[1]{\ensuremath{\text{#1}}} 
  \DeclareMathOperator{\Tr}{Tr} 
\newcommand{\K}{\ix{K}} 
\newcommand{\Ret}{\ix{ret}} 
\newcommand{\Adv}{\ix{adv}} 
\usepackage{bbm}
\begin{document} \title{Luttinger liquid properties of the steady state after a quantum quench}

\author{D.M.\ Kennes}  
\affiliation{Institut f{\"u}r Theorie der Statistischen Physik, RWTH Aachen University 
and JARA---Fundamentals of Future Information
Technology, 52056 Aachen, Germany}

\author{V.\ Meden} 
\affiliation{Institut f{\"u}r Theorie der Statistischen Physik, RWTH Aachen University 
and JARA---Fundamentals of Future Information
Technology, 52056 Aachen, Germany}

\begin{abstract} 

We study the dynamics resulting out of an abrupt change of the two-particle interaction in
two models of closed one-dimensional Fermi systems: (a) the field theoretical 
Tomonaga-Luttinger model and (b) a microscopic lattice model.  
Using a nonperturbative approach which is controlled for small two-particle 
interactions we are able to reach large times allowing us to access the properties of 
the steady state of the lattice model. Comparing those to the exact solution
of the full dynamics in the Tomonaga-Luttinger model we provide evidence for universal 
Luttinger liquid behavior. 

\end{abstract}

\date{\today} 
\maketitle

\section{Introduction}

With the rapid progress in the preparation and measurement techniques for isolated cold  
gases \cite{Bloch08} investigating the fundamental questions of if and how a closed 
quantum many-body system prepared in a nonequilibrium initial state approaches a stationary one  
is within experimental reach. Studying the physics of the steady 
state itself is of particular interest if it is nonthermal,\cite{Polkovnikov11} that is
expectation values of observables differ from those 
computed using a canonical statistical operator with the temperature fixed by the excess 
energy. One-dimensional (1d) interacting Fermi 
systems are promising candidates for realizing such unusual nonequilibrium states as 
in many of those the dynamics is restricted not only by energy conservation 
but by additional conservation laws.\cite{Rigol07} An often studied 
protocol, which we also employ, is an abrupt quench of the 
amplitude $U$ of the two-particle interaction: the system is prepared in the 
canonical ensemble of an initial Hamiltonian $H(U_{\rm i})$, while the time evolution is 
performed with $H(U_{\rm f})$, $U_{\rm f} \neq U_{\rm i}$.  After taking the thermodynamic 
limit local observables might become stationary at long times $t \to \infty$.\cite{Barthel08,Fagotti12} 

In equilibrium the exactly solvable Tomonaga-Luttinger (TL) 
model \cite{Giamarchi03,Schoenhammer05} is the infrared fixed point model under 
a renormalization group (RG) flow of a large class  
of interacting 1d models in their ungapped metallic phase.\cite{Solyom79} The low-energy
physics of models out of this  Luttinger liquid (LL) universality class \cite{Haldane80} 
is given by that of the TL model. For spinless models the fixed point is 
characterized by the velocity $v$ of the elementary excitations and the LL parameter 
$K$ entering exponents of algebraically decaying correlation functions. Both depend 
on the parameters of a given model, in particular on $U$.
 
One of the hallmarks of LLs is their sensitivity towards 
inhomogeneities.  For repulsive interactions with $K<1$ the ground state density response 
function of a LL diverges as \cite{Luther74} $\chi(q,\omega=0) \sim |q- 2 k_{\rm F}|^{2K-2}$, 
with the Fermi momentum $k_{\rm F}$, indicating that even a single weak impurity 
acts as  a strong perturbation. The homogeneous perfect chain fixed point is 
unstable.\cite{Kane92,Andergassen04,Enss05} The system flows towards 
an open chain one with strong consequences for the equilibrium low-energy properties; 
e.g.~the linear conductance vanishes as $G \sim T^{2K^{-1}-2}$ for temperature $T \to 0$.        

We provide evidence that the steady state of a microscopic 1d 
lattice model after an interaction quench is characterized by the same 
power laws as found in the steady state of the TL model 
after a similar quench \cite{Cazalilla06,Uhrig09,Iucci09,Kennes10,Dora11,Rentrop12}
with the $K$ taken for the considered model parameters. As the RG arguments 
leading to this type of LL universality in equilibrium \cite{Solyom79} cannot 
directly be transferred to nonequilibrium \cite{Mitra11,Mitra12,Mitra13,Rentrop12,Iucci09}
this finding is far from obvious. It complements earlier indications of LL universality 
in the time evolution towards the steady state.\cite{Karrasch12} We compute the time evolution 
of the density as a function of the distance from an open boundary as well as that 
of the conductance across a single impurity as a function of $T$ and take 
$t \to \infty$. The dynamics of the TL model is solved exactly 
using standard bosonization.\cite{vonDelft98,Giamarchi03,Schoenhammer05}
To study the time evolution of the lattice model we use an approximate 
functional RG  \cite{Metzner12} based approach which so far was only 
applied to open quantum systems.\cite{Kennes12} 
For small two-particle interactions this technique allows controlled 
access to time scales large enough such that the physics is dominated 
by the steady state. It complements  
calculations using the density-matrix renormalization group (DMRG) which provide `exact' results at small $t$ but abruptly 
become unreliable beyond a characteristic time scale;\cite{Schollwoeck11,Vidal04,White04,Daley04} 
the latter might be smaller than the one on which the steady 
state is reached.\cite{Karrasch12} 
We show that the fixed point structure of a single impurity 
in a nonequilibrium steady-state LL is similar to the one in equilibrium.\cite{Kane92} 

The rest of the paper is structured as follows. In the following section we discuss the exact solution of 
the quench dynamics of the TL model via bosonization. In Sect.~\ref{sect:mm} we introduce the lattice model considered 
and show how its relaxation dynamics can be treated approximately within the functional RG. We then 
compare the prediction of the TL model for observables and correlation functions of the steady state 
with results obtained for the lattice model. Finally, in Sect.~\ref{sect:oq} we briefly discuss the 
relation of our results 
to those obtained by field-theoretical methods and hint towards open questions. 

\section{Tomonaga-Luttinger Model}

The starting point of our investigation of LL universality
is the exact computation of the desired observables and correlation functions 
within the (spinless) TL model. 
Starting out from the 1d electron gas the TL model is obtained by linearizing the single-particle 
dispersion and keeping only the marginal two-particle scattering terms. In contrast to
earlier studies on interaction quenches in the TL model 
\cite{Cazalilla06,Uhrig09,Iucci09,Kennes10,Dora11,Rentrop12,Karrasch12,Mitra11,Hamerla12,Pollmann13,Dora13} 
we consider the one with open boundaries at $x=0$ and $x=L$.\cite{Kane92,Fabrizio95} 
This allows us to distinguish between bulk and boundary LL exponents.\cite{Mattsson97,Meden00}
While in equilibrium the former are quadratic in the two-particle interaction the latter are linear. 
The model is given by 
\begin{eqnarray} 
 H_{\rm TL}  = \sum_{n =1}^{\infty} k_n \left[ v_{\rm F} 
b_n^\dag b_n^{} + \frac{1}{4 \pi} u(k_n) \left(b_n^\dag + b_n^{} \right)^2 \right],
\label{HTL}
\end{eqnarray}
with the Fermi velocity $v_{\rm F}$, $k_n = n \pi/L$, the 
two-particle potential $u(k)$, and bosonic operators $b_n^{(\dag)}$ associated with the
density of the fermions. Employing a Bogoliubov transformation $H_{\rm TL}$ can be written as a 
diagonal quadratic form in eigenmodes with (bosonic) ladder operators $\alpha_n^{(\dag)}$ and 
energy $\omega_n=v_{\rm F} k_n \sqrt{1+u(k_n)/(\pi v_{\rm F})}$.  
To keep the formulas 
transparent we here take the noninteracting canonical statistical operator 
$\rho_{\rm c}^0=\exp[-\beta  H_{\rm TL}(u=0)]/\Tr\{ \exp[-\beta  H_{\rm TL}(u=0)]\}$ as the initial state 
with $\beta=T^{-1}$; in Appendix \ref{ap:bos}
we describe the changes when starting in the canonical state with $u_{\rm i}(k)>0$. 

Using the Bogoliubov transformation and standard bosonization of the field operator  
\cite{vonDelft98,Giamarchi03,Schoenhammer05,Fabrizio95} 
it is straight forward to derive closed analytical expressions for the density $n_t(x)$,
with $x$ being the distance from the boundary, the density of states (DOS) 
$\rho_t(\omega)$, and the density response $\chi_t(q,\omega)$ (see Appendix \ref{ap:bos}). 
The latter two functions can be used to compute the conductance in the limits of small
and large impurities (see below). The expectation values are obtained by  
taking $\mbox{Tr}  \left[ \rho_{\rm c}^0 e^{i H t} O e^{-iHt} \right]$, where $O$ stands 
either for an observable or the operator product defining a correlation function. 
After performing the thermodynamic limit the steady state values follow 
by taking $t \to \infty$; all the above quantities converge and their steady-state 
limits are indicated by dropping the index $t$. We verified that the same 
$t \to \infty$ expectation values can be computed using the statistical operator of a generalized 
Gibbs ensemble (GGE).\cite{Rigol07,Cazalilla06,Iucci09,Kennes10,Rentrop12,Essler12}

At $T=0$, with the initial state given by the noninteracting ground state, 
the steady state reached after the quench is `critical', that is characterized by 
power-law scaling.\cite{Cazalilla06,Uhrig09,Iucci09,Rentrop12}
As a consequence different observables show characteristic power-law behavior with exponents which 
in general are different to the exponents found in 
equilibrium.\cite{Giamarchi03,Schoenhammer05,Haldane80,Solyom79,Luther74,Kane92} They can all be 
expressed in terms of the models LL parameter 
$K=[1+u(0)/(\pi v_{\rm F})]^{-1/2}$ after the quench. The access density $\Delta n(x)=n(x)-\nu$, 
where $\nu$ denotes the filling, for large distances from the boundary $x$ falls off as
\begin{align}
\Delta n^{\rm{eq}}(x)&\sim x^{-K} \sin(2k_{\rm F} x)\\
\Delta n^{\rm{st}}(x)&\sim x^{-(K^2+1)/2}\sin(2k_{\rm F} x)
\end{align} 
where superscripts ${\rm{eq}}$ and ${\rm{st}}$ refer to the ($T=0$) equilibrium or the steady 
state reached after the interaction quench, respectively. Both cases show  damped Friedel oscillations with 
frequency $2k_{\rm F}$. One finds that as the (repulsive) interaction strength is increased and thus 
$K$ becomes smaller (starting at $K=1$ for the noninteracting case), the access density in the presence of a 
boundary in both cases falls off slower than for vanishing two-particle interaction, but with 
different exponents. The difference between ground- and steady-state exponents in the TL model was 
emphasized before considering other observables and correlation functions.\cite{Cazalilla06,Uhrig09,Iucci09,Rentrop12} 

\begingroup
\squeezetable
\begin{table}
\begin{ruledtabular}
\begin{tabular}{cccc}
  observable/correl. funct. &  variable & eq. exp. & steady-state exp. \\ \hline
 access density $\Delta n$ & $x$ & $-K$ & $-(K^2+1)/2$  \\
 local DOS  $\rho$& $\omega$ & $K^{-1} -1 $ &$ (K^{-2}-1)/2 $ \\
 bulk $\chi$ at $\omega=0$ & $q-2 k_{\rm F}$ & $ 2(K-1)$ & $ K^2-1$
\end{tabular}
\end{ruledtabular}
\caption{Equilibrium and steady-state scaling exponents. \label{exponents}} 
\end{table}
\endgroup

Additionally, we consider the frequency resolved local DOS $\rho(\omega)$ 
at small $|\omega|$ and close to the boundary. In the ground state it is suppressed as 
\begin{equation}
\rho^{\rm{eq}}(\omega)\sim |\omega|^{K^{-1}-1}, 
\end{equation}   
which changes to 
\begin{equation}
\label{eq:bla}
\rho^{\rm{st}}(\omega)- \rho^{\rm st} (0) \sim |\omega|^{(K^{-2}-1)/2}
\end{equation} 
in the nonequilibrium steady state. In contrast to the equilibrium case in the steady state
reached after the interaction quench the DOS takes a finite value $\rho^{\rm st} (0)$ at 
$\omega=0$.\cite{Kloeckner13} 
This incomplete suppression is reminiscent of the equilibrium DOS at finite temperatures 
for which the zero frequency value scales as $T^{K^{-1}-1}$.\cite{Mattsson97} 
The exponents of the power law behavior with which the $\omega=0$ spectral weights are reached 
differ between the equilibrium and steady-state situation. Both increase with increasing 
interaction.

Finally, we compute the zero frequency bulk charge susceptibility $\chi(\omega=0)$ for wave vectors 
close to the backscattering condition $q=2k_{\rm{F}}$. The divergence ($K<1$ for repulsive interactions) 
changes from the ground-state result
\begin{equation}
\chi^{\rm{eq}}(\omega=0,q-2k_{\rm{F}})\sim \left(q-2k_{\rm{F}}\right)^{2(K-1)}
\end{equation}
to
\begin{equation}
\chi^{\rm{st}}(\omega=0,q-2k_{\rm{F}})\sim \left(q-2k_{\rm{F}}\right)^{K^2 -1}
\end{equation} 
in the steady state. 
The comparison of the equilibrium and steady state exponents is summarized in Table \ref{exponents}.

In the next section we directly compare 
the decay of the densities Friedel oscillations off the boundary in the steady state of a 
microscopic lattice model with the TL model prediction.

The scaling behavior of the bulk static density response $\chi^{\rm st}(q,\omega=0)$
allows us to make predictions for how the steady state reacts to a single impurity. 
For repulsive interactions $K^2-1 < 0$ and $\chi$ diverges for $q \to 2 k_{\rm F}$. 
As in equilibrium even a weak single impurity strongly disturbs 
the homogenous system. When applying an infinitesimal bias voltage $V$ across the 
impurity the steady-state linear conductance $G=dI^{\rm st}/dV$ (with the stationary 
current $I^{\rm st}$) is expected to scale 
as $G_0-G(T) \sim T^{K^2-1}$, with the constant homogenous chain 
conductance $G_0$.
The power law holds as long as the right hand side stays small, that is for not too 
small $T$. Using the language of equilibrium RG this indicates that the perfect chain 
fixed point is unstable. In contrast, 
the steady-state analog of the open chain one is stable as follows from the 
scaling of the local DOS. Fermis Golden Rule-like arguments lead to a tunneling 
conductance across a weak link connecting two semi-infinite chains which scales as $G \sim T^{K^{-2}-1}$ in a temperature regime which at
the lower end is cut off by the finite DOS at $\omega=0$ [see Eq. \eqref{eq:bla}].
These arguments 
do not rule out intermediate impurity fixed points. 
Provided the concept of LL universality holds for the steady state 
we expect to find those weak and strong impurity scaling laws of $G(T)$ for a lattice 
model with the $K$ of the  model considered. 

\section{Microscopic model}
\label{sect:mm}

We consider the lattice model of spinless fermions
with nearest-neighbor hopping $J$ as well as interaction $U$ and open boundaries 
terminating the $N$-site chain given by  
\begin{eqnarray}
H_{\rm LM}(U)\! =\! \sum_{j=1}^{N-1}  \!\left[- J c_j^{\dag} 
c_{j+1}^{\phantom{\dagger}} + \mbox{H.c.} + U c_j^\dag  c_j^{\phantom{\dagger}}  c_{j+1}^\dag c_{j+1}^{\phantom{\dagger}} \right] 
\label{Hlatticemodel} 
\end{eqnarray}
in standard second quantized notation.
In equilibrium the model is (A) Bethe ansatz solvable and (B) shows universal LL physics 
with $K$ and $v$ exactly known.\cite{Haldane80} It is 
commonly  believed that because of (A) the steady state after an interaction quench 
is described by a GGE but the corresponding statistical operator was so far neither constructed 
nor was a proof of its existence given. Our analysis does not rely on any such assumption. 
When discussing the impurity physics $H_{\rm LM}$ is supplemented by a hopping impurity 
$H_{\rm imp} = h c_{N/2}^\dag  c_{N/2+1}^{\phantom{\dagger}} +  \mbox{H.c.}$ of strength $h \in [0,J]$ 
located in the middle of the chain. 
\subsection{Method}
To compute the time
evolution of the density $n_j$ as well as the conductance $G$
we use an approximate functional RG \cite{Metzner12,Kennes12} based approach. Here, we employ the lowest order truncation scheme in the two-particle interaction. To this order the self-energy acquires a RG flow, which is crucial to capture the impurity physics,\cite{Kane92,Andergassen04,Enss05} while the two-particle vertex remains the bare one. Renormalization of the latter is a higher order effect. The same truncation level was 
earlier shown to capture the equilibrium LL properties of inhomogeneous lattice models 
 including the characteristic power-law scaling, with exponents 
agreeing to the exact ones to leading order in $U$.\cite{Andergassen04,Enss05} . Motivated by the functional RG's success in describing the equilibrium properties of inhomogeneous lattice models, we extend it to tackle the quench dynamics in such systems. 
As the functional RG can directly be applied to the microscopic model, i.e. without the need of mapping it to an effective low-energy field-theory, the information about the high energy modes is not lost and one can hope to find reliable results also for the relaxation at short times (being influenced by the high energy characteristics of the underlying lattice model) as well as the crossover behaviour.
We study the relaxation dynamics and the steady state of a closed many-body system 
described by a lattice model of spinless fermions with nearest neighbor hopping 
and interaction. Compared to the functional RG approach to time evolution 
for open quantum systems introduced in Ref.~\onlinecite{Kennes12}, some minor amendments 
need to be made. Those are outlined next. 

We can treat Hamiltonians of the form
\begin{align}
H&=H_0+H_\ix{int},\\
H_0 &= \sum_{ij} \epsilon_{ij} c_i^\dagger c_j,\\
H_\ix{int}  &= \frac{1}{4} \sum_{ijkl} \bar u_{ijkl} c^\dagger_i
  c^\dagger_j c_l c_k
\end{align}  
written in standard second quantization. Here $\bar u_{ijkl}$ is the antisymmetrized 
two-particle interaction. For our lattice model $H_0=H_{\rm LM}(U=0)$ and $H_\ix{int}=\sum_{j=1}^{N-1}  U c_j^\dag  c_j^{\phantom{\dagger}}  c_{j+1}^\dag c_{j+1}^{\phantom{\dagger}} $ The indices $i,j,\ldots$ stand for the quantum numbers, 
e.g.~the $N$ Wannier states in our lattice model. 
We assume an initial density matrix 
\begin{equation}
\rho^0= \frac{e^{\sum_{ij} \beta_{ij} c_i^\dagger c_j}}{\Tr\left[e^{\sum_{ij} \beta_{ij} c_i^\dagger c_j}\right]}
\end{equation} 
which allows for the application of Wick's theorem. For our calculations we always choose the 
noninteracting canonical statistical operator $\rho^0=\rho^0_{\rm c}=\exp(-\beta H_0)/\Tr[\exp(-\beta H_0)]$. 
We introduce a cutoff in the noninteracting Keldysh \cite{Rammer07} Green functions $g$ 
(as motivated in Ref.~\onlinecite{Kennes12}) by considering
\begin{align}
  \label{eq:gRet}
  g^{\Ret,\Lambda}(t,t') &= - i \Theta(t-t')  e^{-i  \epsilon (t-t') }e^{-i\Lambda(t-t')}\\
    &=\left[ g^{\Adv,\Lambda}(t',t)\right]^\dagger,
  \\
  \label{eq:gK}
  g^{\K,\Lambda}(t,t') &= -i g^{\Ret,\Lambda}(t,0) (1-2 \bar n) g^{\Adv,\Lambda}(0,t'),
\end{align}
with $\epsilon$ being the $N\times N$ matrix with entries $ \epsilon_{ij} $ and $
  \bar n_{ii'} = \Tr \left[ \rho^0  d^\dagger_{i'} d_i \right]$.
The self-energy is obtained by solving a set of coupled differential flow equations 
\begin{align}
&\partial_\Lambda \Sigma^{\text{ret},\Lambda}_{i_1 i_{1'}}(t',t)=\partial_\Lambda \Sigma^{\text{adv},\Lambda}_{i_1 i_{1'}}(t',t)\notag\\&=-\sum\limits_{i_2,i_{2}'}S^{\text{K},\Lambda}_{i_2' i_2}(t,t)\left(-i \bar u_{i_1 i_2 i_1' i_2'}(t)\right)\delta(t'-t),\label{eq:FlowRet}\\
&\partial_\Lambda \Sigma^{\text{K,}\Lambda}=0 , 
\end{align}
with the initial conditions at $\Lambda= \infty$ 
\begin{align}
& \Sigma^{\ix{ret}, \Lambda=\infty}_{i'i}(t',t) = \frac{1}{2}
  \delta(t-t') \sum_j \bar u_{i'jij},\\
&  \Sigma^{\ix{K},\Lambda=\infty}_{i'i}(t',t) = 0 . 
\end{align}
The right hand sides of the flow equations contain  
\begin{equation}
\begin{split}
 S^{\K,\Lambda} &=  \partial^*_\Lambda  G^{\K,\Lambda},
\end{split}
\end{equation}
with the full cutoff dependent Keldysh component of the Green function 
\begin{equation}
G^{\K,\Lambda}(t,t')= -i G^{\Ret,\Lambda}(t,0) (1-2 \bar n) G^{\Adv,\Lambda}(0,t') .
\label{eq:GKL}
\end{equation}
The star differential operator $\partial^*_\Lambda$ acts only on the free 
Green function $ g^{\Ret/\Adv,\Lambda} $, not on $ \Sigma^\Lambda $, in the Dyson series expansion 
$  G^{\Ret/\Adv,\Lambda}=  g^{\Ret/\Adv,\Lambda}+ g^{\Ret/\Adv,\Lambda}\Sigma^\Lambda  g^{\Ret/\Adv,\Lambda}+\dots$ 
used to calculate $ G^{\Ret/\Adv,\Lambda} $. 
An approximation to the self-energy of the cutoff-free problem is obtained 
at $\Lambda=0$.
How to efficiently evaluate Eq. \eqref{eq:GKL} is summarized in Appendix \ref{ap:num_im}.

An approximation for the occupancy of site $j$ can directly be obtained from the Keldysh Green function 
at the end of the RG flow as
\begin{equation}
   n_j(t) = \frac{1}{2} - \frac{i}{2} G^{\K,\Lambda=0}_{jj}(t,t).
\label{occu}
\end{equation} 

To calculate the current flowing from the left to the right half of the lattice in our 
microscopic model of interacting spinless fermions, we need to determine
\begin{equation}
I(t) =-\frac{d}{dt}\left\langle N_L(t)\right\rangle_{\rho^0},
\end{equation}  
where $N_L$ is defined as $N_L(t)=\sum_{j=1}^{N/2} n_j(t)$, with 
$n_j =c_j^\dagger c_j$ being the occupancy operator of site $j$ 
and $\left\langle  \ldots \right\rangle_{\rho^0}$ denotes the expectation value with respect to 
$\rho^0$.\cite{Meir92} For simplicity we assume an even number of lattice sites $N$. Furthermore, we use 
\begin{equation}
I(t)=-i\left\langle[H, N_L](t)\right\rangle_{\rho^0}=(h-J)G^<_{N/2 N/2+1}(t,t)+{\rm c.c}.\label{eq:It}
\end{equation}
with $G^<_{N/2 N/2+1}(t,t)$ being the equal-time lesser Green function of the interacting 
system.\cite{Rammer07} The functional RG method used here provides  an approximation 
for this given by 
\begin{equation}
 G^{<,\Lambda=0}_{i j}(t,t) = \frac{1}{2}\left[i - G^{\K,\Lambda=0}_{ij}(t,t)\right].
\end{equation} 
Therefore, plugging this lesser Green function into Eq.~\eqref{eq:It} allows to 
compute an approximation to the current and from this the conductance by numerical differentiation.

\begin{figure}[t]
\includegraphics[width=0.95\linewidth,clip]{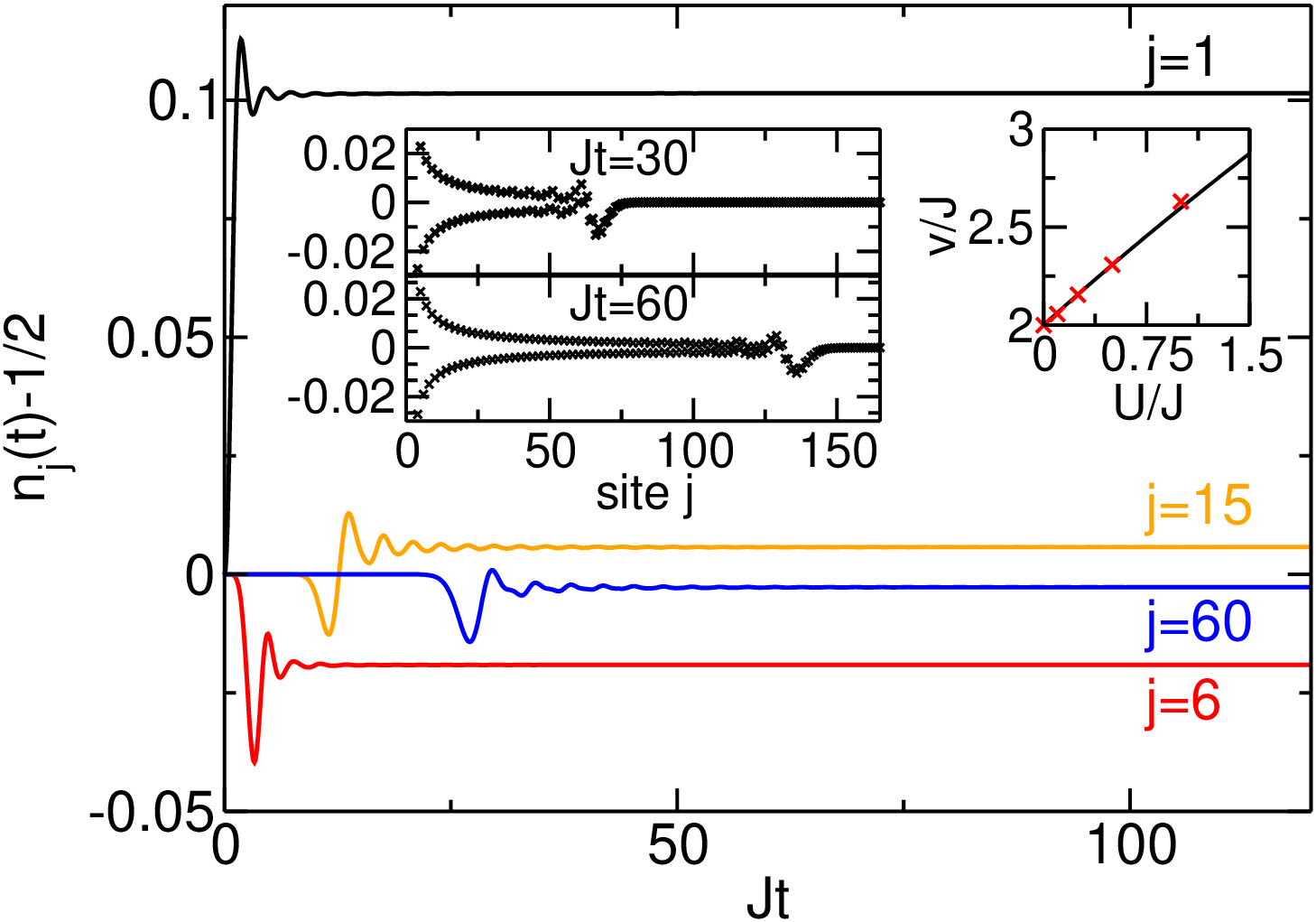}\vspace*{-0.2cm}
\caption{(Color online) Functional RG data for the time evolution of $n_j(t)-\nu$ at half 
filling $\nu=1/2$ after a quench in the interaction amplitude from $U/J=0$ to $U/J=0.5$ for 
$N=10^3$, $T=0$, and different sites $j$. Left inset: Friedel oscillations 
induced by the boundary and the propagation of a main signal from the boundary 
to $j \approx vt$ for two values of $t$. Right inset: Velocity of the 
main signal for different $U$ (symbols). The exact Bethe ansatz $v$ (line) 
is in excellent agreement with our data.}
\label{fig1}
\end{figure}

\subsection{Results}
In Fig.~\ref{fig1} we show the access density $n_j(t)-\nu$, with the filling $\nu$, for fixed $j$ 
starting out of the noninteracting impurity free ground state ($T=0$). 
We can reach times of the order of a few $10^2/J$
which has to be contrasted to the DMRG approach which becomes unreliable for times of the 
order of $10/J$.\cite{Karrasch12} As shown in the main plot 
and the left inset a signal originating from the left boundary travels through the system. 
A similar one is generated at the right one. For a spatial region in which the left signal 
passed through and the right one did not enter yet the density becomes stationary. The physics 
for $t \approx 10^2/J$ and $j$ up 
to ${\mathcal O}(10^2)$ is thus dominated by the steady state and does barely suffer from finite 
size effects. The two signals propagate with the LL 
velocity $v$,\cite{Calabrese07,Langer09,Ganahl12} with 
our method providing an excellent approximation to the exact $v$ (right inset of 
Fig.~\ref{fig1}).    
\begin{figure}[t]
\includegraphics[width=0.95\linewidth,clip]{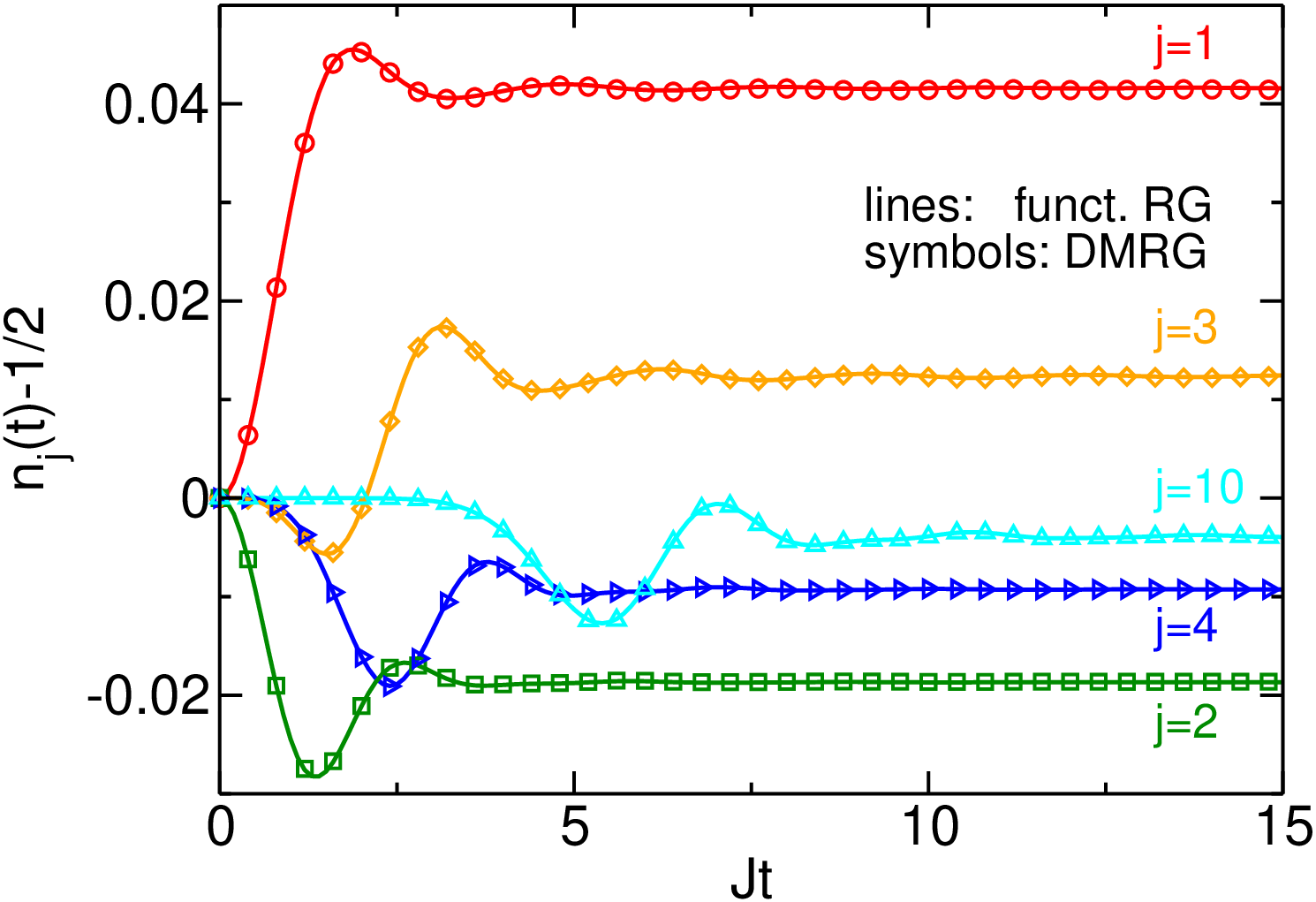}\vspace*{-0.2cm}
\caption{(Color online) Comparison of functional RG and DMRG data for the time evolution of $n_j(t)-\nu$ at half 
filling $\nu=1/2$ after a quench in the interaction amplitude from $U/J=0$ to $U/J=0.2$ for 
$N=10^2$, $T=0$, and different sites $j$. Functional RG and the numerically exact DMRG are in excellent agreement 
for times reachable by the DMRG calculation.}
\label{fig1_DMRG}
\end{figure}

In Fig.~\ref{fig1_DMRG} we compare the functional RG data to numerically exact DMRG results for the time evolution
after a quench in the interaction amplitude from $U/J=0$ to $U/J=0.2$. This value of $U$ is of the same order as 
taken for all further analysis. On the scale of the plot the two datasets are indistinguishable for all sites $j$ 
and all times reachable with DMRG.\cite{Schollwoeck11,Vidal04,White04,Daley04} Within our DMRG approach to prepare 
the noninteracting ground state an iterative single site algorithm in matrix product state formulation was employed. 
The resulting wavefunction was than subjected to a real time evolution using a fourth order Suzuki-Trotter 
decomposition ($J \Delta t=0.2$) ensuring that the discarded weight stays below a certain $\epsilon$ (different 
$\epsilon$ ranging from $10^{-5}$ to $10^{-8}$ were tested to yield coinciding results). The technical details are 
described in length in Ref. \onlinecite{Schollwoeck11}. The excellent agreement of the results obtained by both 
methods does not only strengthen our confidence in the functional RG approach for the following steady state 
analysis, but also shows that indeed within functional RG we correctly incorporate also the high energy physics 
of the underlying lattice model, which is crucial for short times. The numerically exact solution of the time 
evolution within DMRG can however not be pushed to times large enough to allow for the sensitive analysis of 
power-law scaling conducted in the following with our functional RG approach. 

\begin{figure}[t]
\includegraphics[width=0.95\linewidth,clip]{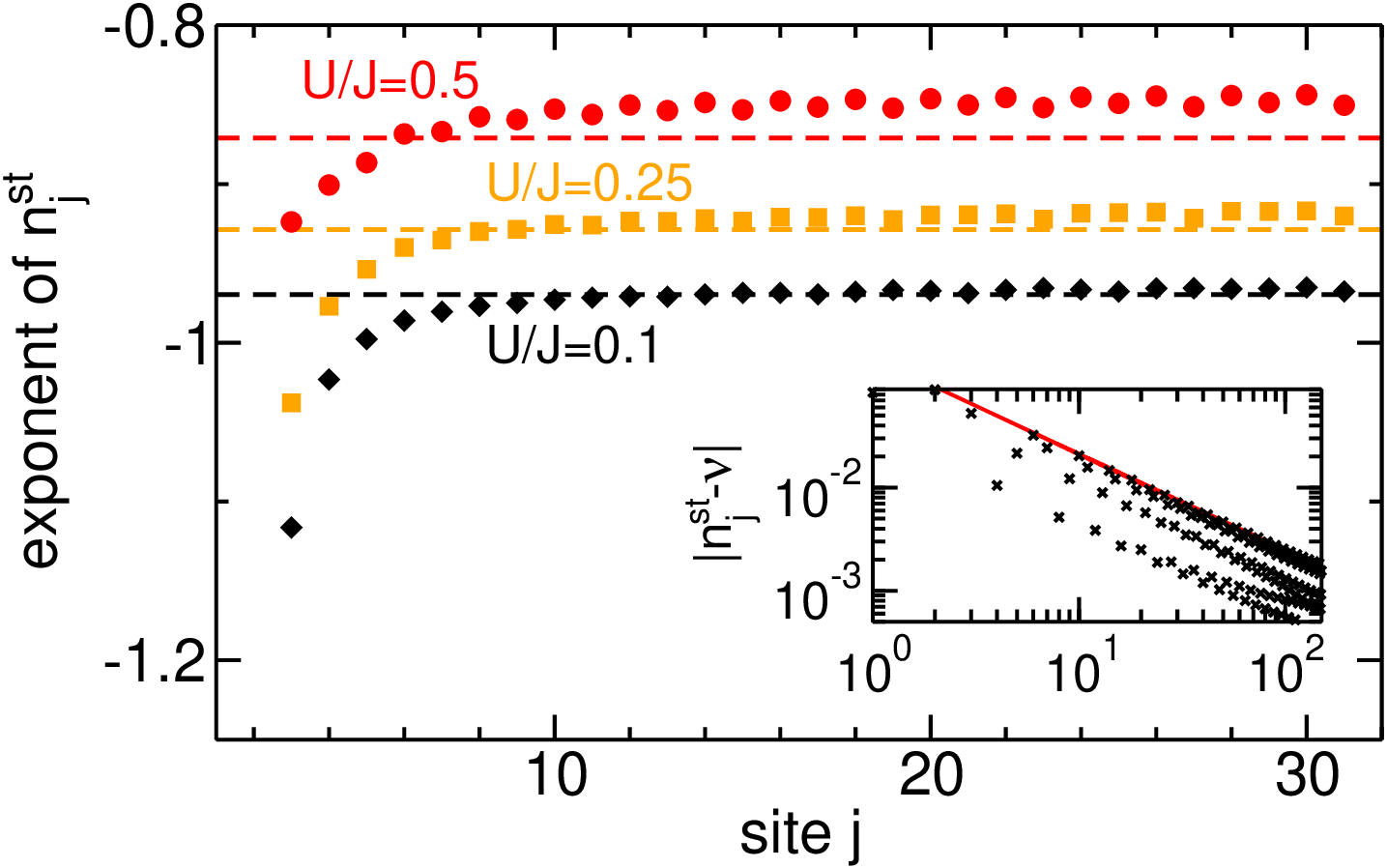}\vspace*{-0.2cm}
\caption{(Color online) Functional RG data for the effective exponent of the 
Friedel oscillations of the 
steady-state density $n_j^{\rm st}$ for $\nu=1/2$, $N=10^3$, and $T=0$ 
determined by taking the log-derivative. The predictions from the TL model 
with the exact lattice model $K$ (dashed lines) are consistent with our results. 
The  inset shows $|n_j^{\rm st}-\nu|$
at $\nu=0.375$ and $U/J=0.25$ (symbols). The line is the TL model prediction.}
\label{fig2}
\end{figure}

In analogy to the ground state density \cite{Andergassen04} the stationary 
one $n_j^{\rm st}$ shows  Friedel oscillations with frequency $2 k_{\rm F}$ 
(see the $j < 80$ region of the lower part of the left inset of Fig.~\ref{fig1}). We next analyze their 
decay. Figure \ref{fig2} shows  the log-derivative of $|n_j^{\rm st}-1/2|$, that is an effective 
exponent. The dashed lines is the prediction from the TL model of Table \ref{exponents} with 
the exact lattice model $K$. Our data are consistent with a power-law decay and the TL model 
exponent. This finding is our first indication of LL universality of the steady state. 
The differences between the exact exponent and our result is of order $(U/J)^2$.
On the right hand side of our RG flow equations we do not fully 
capture terms $\sim U^2$ and thus control exponents only to order $U/J$. 
The discussed behavior is not restricted to the case of half filling. 
The inset of Fig.~\ref{fig2} shows $|n_j^{\rm st}-\nu|$ for $\nu=0.375$ on a log-log scale and 
the corresponding TL model prediction as the envelope. 

To compute the steady-state linear conductance of the lattice model we take the canonical 
density matrix  $\rho_{\rm c}^0$ (with $T>0$) corresponding to  
$H_{\rm LM}(U=0)+ H_{\rm imp}$ as the initial
state. The time evolution is performed with $H_{\rm LM}(U>0) + H_{\rm imp}$ supplemented by onsite 
energies $V/2$ ($-V/2$) for all $j \leq N/2$ ($j>N/2$). 
The current $I(t)$ across the impurity bond is computed. 
Following the same reasoning as for the density $I$ becomes stationary for 
$t$ of the order of $10^2/J$. We take $V$ to be the smallest energy scale of 
the problem (typically $V= 10^{-3} J$) to ensure that we are in the linear 
regime $I^{\rm st} = G V$.  For $T \gtrapprox J$ we find $G(T) \sim T^{-1}$,\cite{Enss05} 
which is a band effect 
(see the inset of Fig.~\ref{fig4}). The universal scaling of the conductance 
discussed in the last section can only be expected for $T \ll J$. 

\begin{figure}[t]
\includegraphics[width=0.95\linewidth,clip]{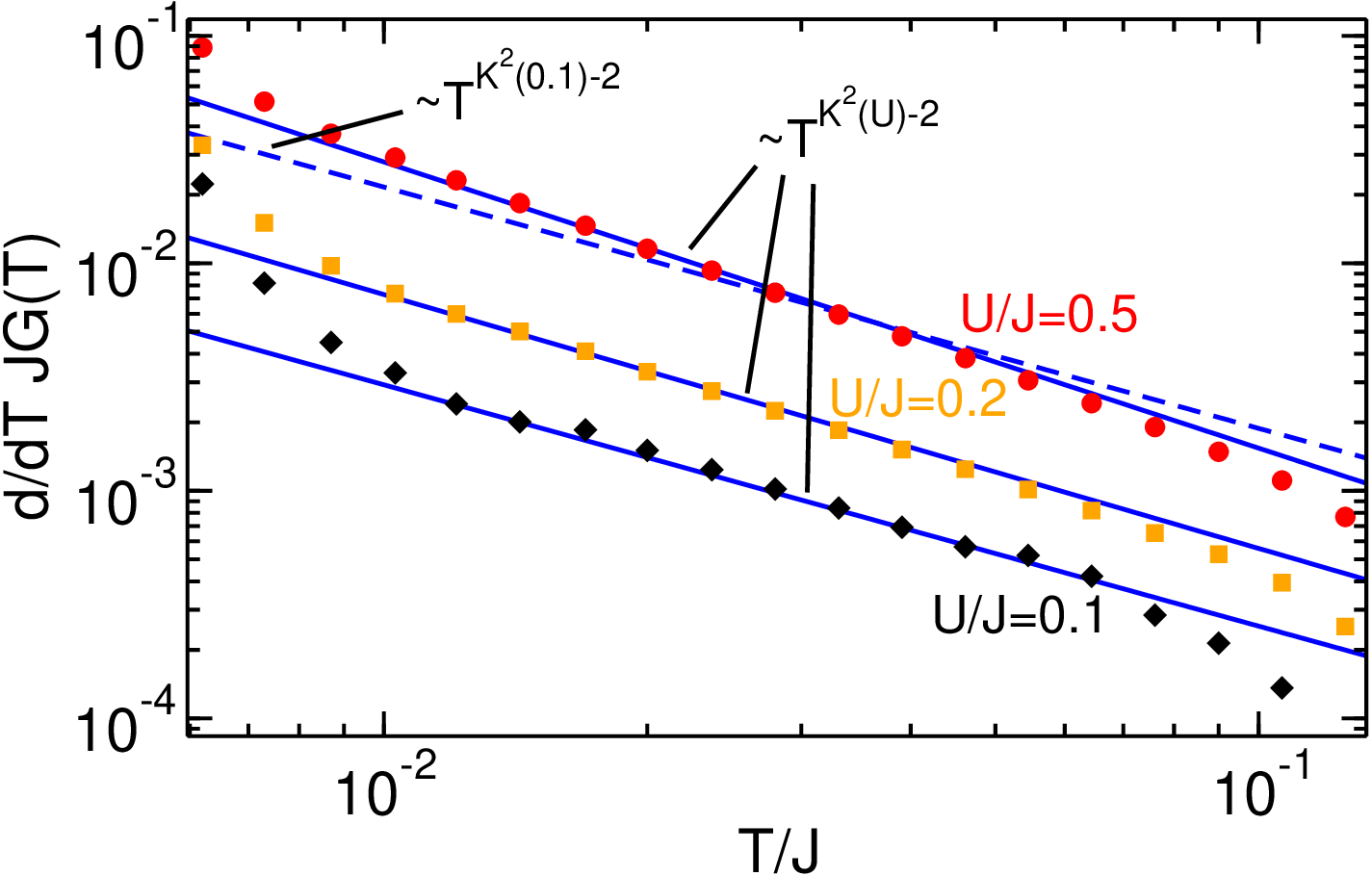}\vspace*{-0.2cm}
\caption{(Color online) Functional RG data for the temperature derivative of the steady-state 
linear conductance of our lattice model for a weak impurity $h/J=0.05$, filling 
$\nu=1/2$, and $N=10^3$ (symbols). 
The prediction from the TL model with the exact lattice model $K$ is 
shown as the solid lines. To emphasize the sizable differences in the exponents 
we have added a power law with the $U/J=0.1$ exponent as the dashed line 
to the $U/J=0.5$ data.}
\label{fig3}
\end{figure}
    
We first analyze the case of weak impurities. To eliminate the 
constant $G_0$ we take the derivative of $G$ with respect to $T$. Based on our above
considerations we expect to find $d G/dT \sim T^{K^2-2}$; see Fig.~\ref{fig3}. Over 
roughly one order of magnitude the functional RG data follow the TL model prediction with 
the exact $K$ of our lattice model. The deviations for $T/J > 0.1$ indicate 
the crossover to the nonuniversal $G(T) \sim T^{-1}$ regime. The ones 
for $T/J < 0.01$ have two reasons. As discussed above the scaling only holds as long 
as $T$ does not become too small. Furthermore, the energy level spacing 
$\delta_N = v_{\rm F}/N$ ($=2 \cdot 10^{-3} J$ for the parameters of the plot) is an 
energy scale of the problem which cuts off any universal behavior.\cite{Andergassen04,Enss05} 
This is an artefact of our treatment of finite systems. 
For small $h$ and $T/J \in [0.1,0.005]$, $G_0-G(T) \ll 1$. Our analysis thus 
requires very accurate data. To minimize the error due to small residual oscillations 
of $I(t)$ present even for $t$ of the order of $10^2/J$ we averaged the data at large 
$t$ over a small time interval.        

\begin{figure}[t]
\includegraphics[width=0.95\linewidth,clip]{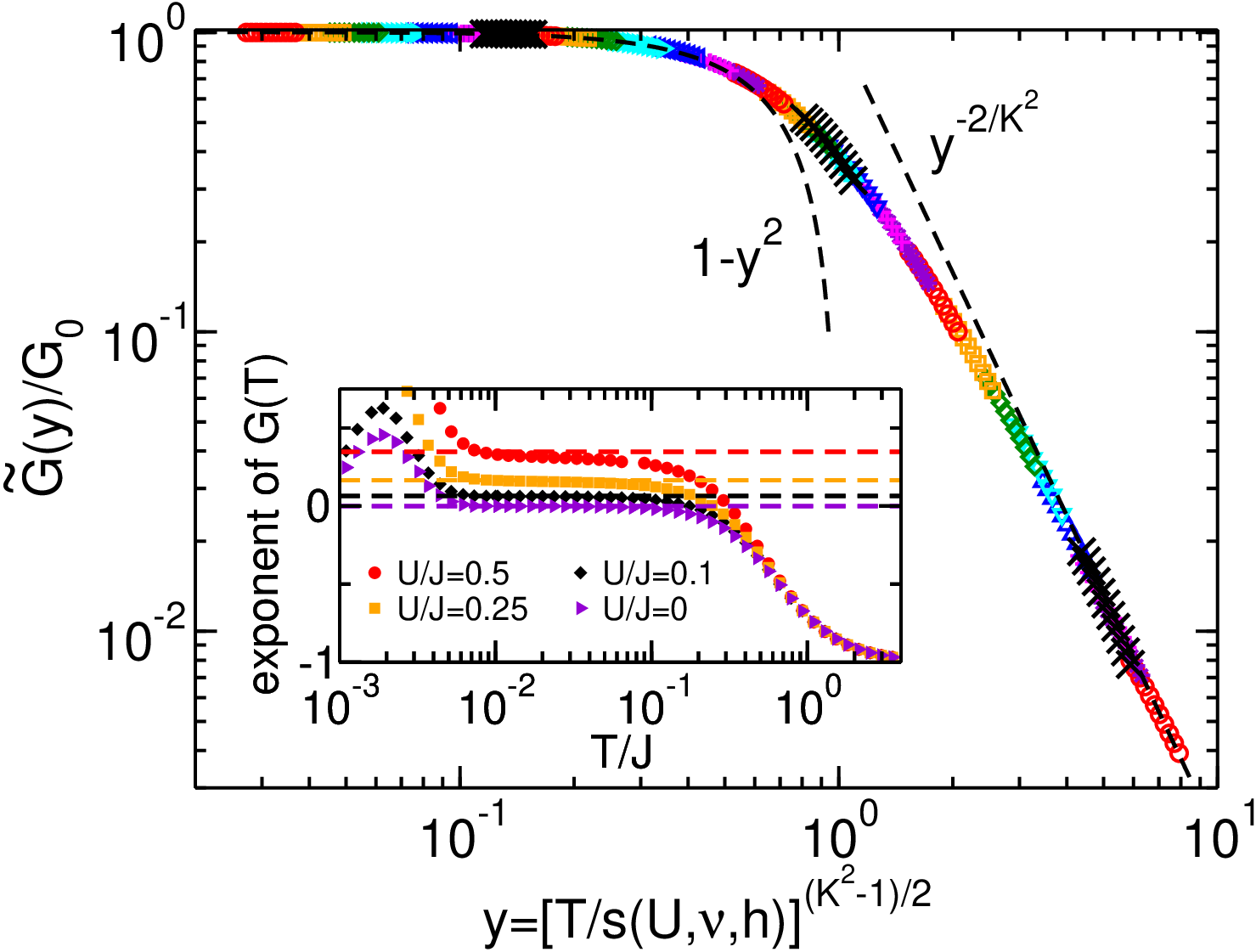}\vspace*{-0.2cm}
\caption{(Color online) Functional RG data for the one-parameter scaling 
of  $\tilde G(y)$ of the lattice model for $U/J=0.5$, $\nu=1/2$, and $N=10^3$. 
Different symbols stand for different $h$ increasing from left to right.
No fixed point in between the perfect and the steady-state analog of the 
open chain ones exist. 
The crosses were computed for $U/J=0.85$, $\nu=1/4$ giving the same
$K$. The scaling function thus depends on $U$ and $\nu$ only via $K$.\cite{Kane92} 
The prediction of the TL model for $y\to 0$ and $y\to \infty$ with the 
exact lattice model $K$ is shown as dashed lines. 
Inset: the effective exponent of $G(T)$ for a strong impurity 
$h/J=0.9$ determined by a log-derivative. Dashed lines show the prediction of the TL model 
with the exact lattice model 
$K$.}
\label{fig4}
\end{figure}

In the inset of Fig.~\ref{fig4} we present our results for $G(T)$ across a strong 
impurity. Even without any $t$ averaging our data are accurate enough 
such that the log-derivative, that is the effective exponent, gives a smooth curve. The 
data clearly show the crossover from the nonuniversal $T^{-1}$ behavior at large 
$T$ to the TL model prediction $T^{K^{-2}-1}$ at low ones. For $T \to 0$ the scaling is 
cut off by both the finite size scale $\delta_N$ as well as the finite
DOS at $\omega=0$ [see Eq.~(\ref{eq:bla})]. In the limits of strong and weak impurities our results for 
the linear conductance of the lattice model thus agree to the TL 
model prediction providing the second indication of LL universality 
of the steady state. 

We finally show that in the steady state of the lattice model no fixed point in 
between the perfect and the steady-state analog of the open chain ones exist. To this end we 
compute $G(T)$ for a variety of $h$ at fixed $U$ and $\nu$. Using a 
one-parameter scaling ansatz of the form $G(T) = \tilde G(y)$,\cite{Kane92,Enss05} 
with $y=(T/s)^{(K^2-1)/2}$ and the nonuniversal scale 
$s(U,\nu,h)$, all data can be collapsed on a single curve continuously connecting the weak 
($y \to 0$)  and strong ($y \to \infty$) impurity 
fixed points; see the main plot of Fig.~\ref{fig4}. 

\section{Open questions}   
\label{sect:oq}

We provided evidence that the steady state of an interacting 1d 
lattice model after a quantum quench shows LL universality. Our analysis relies on 
the functional RG approach in its lowest-order truncation,\cite{Metzner12,Kennes12} 
which is sufficient to obtain LL power laws with $U$-dependent exponents. An obvious first 
question arising is if higher order terms in $U/J$ might change this picture. In fact, 
a series of RG studies of the field-theoretical TL model complemented by  
`perturbations' indicates, 
that power-law scaling is destroyed on long times by certain such terms.\cite{Mitra11,Mitra12,Mitra13} 
In nonequilibrium it is not established if and how the field theory studied 
in those papers is related to microscopic lattice models considered by us.\cite{Mitra13a} 
Currently, the results of Refs.~\onlinecite{Mitra11,Mitra12,Mitra13} and our findings should thus 
be viewed as complementary and not contradicting. We emphasize that the notion of LL universality 
involves lattice models and not only field theories.\cite{Haldane80} The numerically  `exact' 
results of Ref.~\onlinecite{Karrasch12} for the time-evolution towards the steady state are 
consistent with our findings. For the $U/J \leq 0.5$ 
considered by us the corrections of order $(U/J)^2$ are small. Even if they would 
destroy the LL scaling on very large time scales, we expect that remnants of the predicted 
LL steady state can be found up to this scale.  The second apparent open question is if and 
how the picture changes if a lattice model is considered which is not Bethe ansatz 
solvable (`nonintegrable'). We here merely note that for the time dependence 
indications of universal LL power laws were found even for such models.\cite{Karrasch12}

\section*{Acknowledgments}
We thank S.~Andergassen, E. Dalla Torre,
C.~Karrasch, A.~Mitra, K.~Sch\"onhammer, D.~Schuricht, and G.~Uhrig for discussions as well as 
S.~Bl\"ugel and the J\"ulich 
Supercomputing Centre for access to the JuDGE GPU cluster. This work was supported by 
the DFG via FOR 723. 
\appendix
\setcounter{equation}{0}
\section{Bosonization}
\label{ap:bos} 
To compute observables and correlation functions in the steady state 
of the TL model  with open boundaries  after an 
interaction quench we use `open boundary bosonization' for the 
Hamiltonian and the field operator.\cite{Fabrizio95,Mattsson97,Meden00,Grap09}
We are interested in the scaling behavior with all energy scales send to zero and
all length scales send to infinity. Thus subtleties resulting out of the momentum 
dependence of the two-particle potential $u(k)$ become 
irrelevant \cite{Meden99,Rentrop12} and the ultraviolet regularization can 
be implemented at will. To illustrate the procedure we consider the density $n_t(x)$. 
We first study the quench from $u_{\rm i}(k) = 0$ to $u_{\rm f}(k) =  u(k)$. 
For simplicity we focus on temperature $T=0$. The density is given by the Green function 
\begin{eqnarray*}
i G_t(x,x) = \left< \mbox{vac}(b) \right| e^{i H_{\rm TL} t} \psi^\dag(x) \psi(x)^{}  e^{- i H_{\rm TL} t}
 \left| \mbox{vac}(b) \right> ,
\end{eqnarray*}   
with the noninteracting ground state $ \left| \mbox{vac}(b) \right> $ which corresponds to the vacuum 
with respect to the $b$'s [see Eq. \eqref{HTL}]. The fields $\psi^{(\dag)}(x)$ 
contain the open boundary conditions. Using auxiliary fields $\tilde \psi^{(\dag)}(x)$ which are 
identical to the ones obtained for periodic boundary conditions \cite{vonDelft98,Giamarchi03,Schoenhammer05} 
and are e.g.~given in Eqs.~(18)-(20) of Ref.~\onlinecite{Rentrop12}, the Green function reads
\begin{eqnarray*}
i G_t(x,x) & \!   =  \!\! & \frac{1}{2} \! \left[ \left< \mbox{vac}(b) \right| e^{i H_{\rm TL} t} \tilde \psi^\dag(x) \tilde 
\psi(x)^{}  e^{- i H_{\rm TL} t}  \left| \mbox{vac}(b) \right>  \right. \\ 
&& + \left< \mbox{vac}(b) \right| e^{i H_{\rm TL} t} \tilde \psi^\dag(-x) \tilde 
\psi(-x)^{}  e^{- i H_{\rm TL} t}  \left| \mbox{vac}(b) \right>  \\ 
&&  - \left< \mbox{vac}(b) \right| e^{i H_{\rm TL} t} \tilde \psi^\dag(x) \tilde 
 \psi(-x)^{}  e^{- i H_{\rm TL} t}  \left| \mbox{vac}(b) \right>  \\ 
&& \left. - \left< \mbox{vac}(b) \right| e^{i H_{\rm TL} t} \tilde \psi^\dag(-x) \tilde 
 \psi(x)^{}  e^{- i H_{\rm TL} t}  \left| \mbox{vac}(b) \right> \right] .
\end{eqnarray*} 
Those expectation values can be computed following the usual 
steps  \cite{vonDelft98,Giamarchi03,Schoenhammer05} which involve the multiple use of the 
Bogoliubov transformation $b_n^{} = c(k_n) \alpha_n^{} + s(k_n) \alpha_n^\dag$  and the 
Baker-Campbell-Hausdorff relation. The coefficients $c(k_n)$ and $s(k_n)$ depend on 
the two-particle potential $u(k_n)$ and are e.g.~given in Eq.~(9) 
of Ref.~\onlinecite{Rentrop12}.  The first two terms of the Green function 
provide the homogenous background density while the 
latter two oscillate in space with frequency $2 k_{\rm F}$---they contain the Friedel oscillations induced 
by the boundaries. After taking the thermodynamic limit and the limit $t \to \infty$ we find for 
the steady-state access density 
\begin{eqnarray}
\Delta n^{\rm st} (x) \sim x^{-[c^2(0) + s^2(0)] [c(0) + s(0)]} \, \cos{(2 k_{\rm F} x)}   .  
\label{friedel}
\end{eqnarray}
At zero momentum the coefficients of the Bogoliubov transformation can be expressed in terms 
of the models LL parameter $K$ given in the main text as 
\begin{eqnarray}
\mbox{} \hspace{-.4cm} s^2(0)=\frac{1}{4} (K+K^{-1} -2) , \; c^2(0)=\frac{1}{4}(K+K^{-1} +2) .
\label{coeffK}
\end{eqnarray} 
Using those relations the exponent of Eq.~(\ref{friedel}) can be written as $-(K^2+1)/2$ 
(see Table I of the main text). The other scaling exponents of the last column of Table I of the main 
text can be obtained in a similar fashion.

In any (equilibrium) LL the LL parameter is given by 
\begin{eqnarray}
K=1-\tilde U 
+ {\mathcal O}(\tilde U^2) , 
\label{expansion}
\end{eqnarray}
where $\tilde U$ is a dimensionless measure for the interaction strength; 
e.g.~$\tilde U = u(0)/(2 \pi v_{\rm F})$ for the TL model and 
$\tilde U=U [1-\cos(2k_{\rm F})]/[2 \pi J \sin(k_{\rm F})]$ 
for our lattice model at filling $\nu=k_{\rm F}/\pi$. 
Using this expansion it is evident that all scaling exponents discussed by us 
(see Table I of the main text) have a leading order contribution in $\tilde U$. This
is crucial as within our approximate treatment of the lattice model we control exponents 
only to leading order.\cite{Metzner12}    

We next briefly discuss the case when starting in the ground state of $H_{\rm TL}$ with 
$u_{\rm i}(k) > 0$ and performing the time evolution with a different interaction 
$u_{\rm f}(k) > 0$; all quantities depending on the interaction strength acquire indices i or f. 
We have to consider two Bogoliubov transformations [corresponding to the transformations 
from zero interaction to $u_{\rm i/f}(k)$] and two sets of eigenmode ladder operators. 
The initial state is the vacuum with respect 
to one of those while the Hamiltonian with which the time evolution is performed is a 
diagonal quadratic form in the other one. Repeatedly applying the Bogoliubov 
transformations and the Baker-Campbell-Hausdorff relation gives for the 
steady-state access density
\begin{eqnarray}
\Delta n^{\rm st} (x) \sim x^{-\gamma} \, \cos{(2 k_{\rm F} x)}   ,  
\label{friedel_a}
\end{eqnarray}
with the scaling exponent
\begin{eqnarray}
\gamma =  K_{\rm f}   && \left[ 1+ \frac{1}{8} (K_{\rm i} +K_{\rm i}^{-1} -2 )  
(K_{\rm f} +K_{\rm f}^{-1} +2 ) \right. \nonumber \\ 
&& -   \frac{1}{4} (K_{\rm i} - K_{\rm i}^{-1})  
(K_{\rm f} - K_{\rm f}^{-1}) \nonumber \\ && \left. + \frac{1}{8} (K_{\rm i} +K_{\rm i}^{-1} +2 )  
(K_{\rm f} +K_{\rm f}^{-1} -2 ) \right] .
\label{gammaexp}
\end{eqnarray} 
For $K_{\rm i}=1$, that is if we start in the noninteracting ground state,  
it becomes equal to $(K_{\rm f}^2+1)/2$ (see Table I of the main text). Using the 
expansion Eq.~(\ref{expansion}) it is easy to see that $\gamma$ and $(K_{\rm f}^2+1)/2$
agree to leading order in the two-particle interaction, that is $U_{\rm i}$ only contributes 
to order $U_{\rm i}^2$ and $U_{\rm f} U_{\rm i}$ or higher. To leading order in the interaction 
strength the scaling exponent of the density is thus exclusively given by $U_{\rm f}$. 
The same holds for the other exponents considered by us (see Table I of the main text). 
In our computations for the lattice model we control exponents only to leading order which 
explains why in the main text we exclusively consider quenches out of the noninteracting 
ground state.

\section{Numerical implementation of the functional RG}

\label{ap:num_im}
We can calculate $G^{\K,\Lambda}(t,t')$ very efficiently by using an iterative procedure. 
First, we discretize time in steps such that during one small step $\Delta t$ the time dependent 
self-energy can be set constant. For our results shown in the main text we made sure 
that $\Delta t$ is always chosen small enough such that further reducing it does not lead to 
any changes visible on the scale of the respective plots. We use 
\begin{align}
\label{eq:multiformula}
  G^{\Ret,\Lambda}(t,t') &G^{\Ret,\Lambda}(t',t'') \notag\\&= - i \Theta(t-t')
  \Theta(t'-t'') G^{\Ret,\Lambda}(t,t'').\\
  G^{\Ret,\Lambda}(t,t') &= \left[ G^{\Adv,\Lambda}(t',t)\right]^\dagger,
\end{align}
to write $G^{\Ret}(t,t')$ as a product of Green functions\cite{Kennes12}
\begin{align}
  \label{eq:GRet}
  G^{\Ret,\Lambda}(t_1+\Delta t,&t_1) &= - i  e^{-i  [\epsilon+\bar\Sigma^{\Ret,\Lambda}(t_1)] \Delta t }e^{-i\Lambda(t-t')},
\end{align}
where $\bar\Sigma^{\Ret,\Lambda}(t_1)$ is the self-energy time averaged over the interval 
$(t_1,t_1+\Delta t)$. The interacting Keldysh Green function $G^{\K,\Lambda}(t,t)$ can 
then be found iteratively employing
\begin{align}
\mbox{} \hspace{-.2cm} G^{\K,\Lambda}&(t+\Delta t,t+\Delta t)\notag\\&= G^{\Ret,\Lambda}(t+\Delta t,t)G^{\K,\Lambda}(t,t)
G^{\Ret,\Lambda}(t,t+\Delta t),\\
\mbox{} \hspace{-.2cm} G^{\K,\Lambda}&(0,0)=-i(1-2 \bar n).
\end{align}
In every time step two matrix exponentials of $N \times N$ matrices have to be performed and multiplied 
with the Keldysh component of the Green function of the previous one. This renders the problem a natural 
candidate for graphics processing unit (GPU) supported algorithms. We  use such to compute the 
results shown in the main text. The number $N_t$ of time steps needed to obtain sufficient accuracy (and 
resolution) as well as to reach times which are large enough such that the physics is dominated 
by the steady state enters the number  of 
equations to be solve. For the Hamiltonian considered in the main text 
we solve sets of $(3N-2) N_t$ coupled differential equations. Due to the 
nearest neighbor structure of the interaction $3N-2$ components of the self energy flow 
for each of the $N_t$ time steps. Typical numbers considered are $N=1000$ lattice sites and 
$N_t = 1600$ time steps. 
{}

\end{document}